\documentclass[5p]{elsarticle}

\usepackage{amsmath}
\DeclareMathOperator\erfc{erfc}
\usepackage{amssymb}
\usepackage{mathtools}
\usepackage{siunitx}
\DeclareSIUnit{\atpercent}{at.\%}
\usepackage{url}
\usepackage{lineno,hyperref}
\hypersetup{
pdfborder={0 0 0},
breaklinks=true,}
\usepackage{xcolor}
\definecolor{Cr}{RGB}{83,83,83}
\definecolor{Co}{RGB}{0,78,138}
\definecolor{Fe}{RGB}{127,171,22}
\definecolor{Ni}{RGB}{204,76,3}
\definecolor{Mn}{RGB}{185,15,34}
\definecolor{tud11c}{RGB}{97,28,115}

\usepackage{multirow}
\usepackage{graphicx}
\usepackage{rotating}
\usepackage{booktabs}
\usepackage{pgfplots}
\usepackage{pgfplotstable}
\pgfkeys{/pgf/number format/.cd,1000 sep={}}
\pgfplotsset{compat=newest}
\journal{Acta Materialia}

\begin{document}


  \begin{frontmatter}
  
  \title{
  Experimental and theoretical study of tracer diffusion in a series 
  of (CoCrFeMn)$_{100-x}$Ni$_x$ alloys}
  
  \author[muenster]{Josua Kottke\corref{mycorrespondingauthor}}
  \cortext[mycorrespondingauthor]{Corresponding author}
  \ead{josua.kottke@uni-muenster.de}
  
  \author[darmstadt]{Daniel Utt\corref{mycorrespondingauthor}}
  \ead{utt@mm.tu-darmstadt.de}
  
  \author[paris]{Mathilde Laurent-Brocq}
  \author[muenster]{Adnan Fareed}
  \author[muenster]{Daniel Gaertner}
  \author[paris]{Lo\"ic Perri\`ere}
  \author[krakow]{\L{}ukasz Rogal}
  \author[darmstadt]{Alexander Stukowski}
  \author[darmstadt]{Karsten Albe}
  \author[muenster]{Sergiy V. Divinski}
  \author[muenster]{Gerhard Wilde}
  
  \address[muenster]{Institute of Materials Physics, University of M\"{u}nster, 48149 M\"{u}nster, Germany}
  \address[darmstadt]{Fachgebiet Materialmodellierung, Institut f\"ur Materialwissenschaft, TU Darmstadt, 64287 Darmstadt, Germany}
  \address[paris]{Universit\'{e} Paris-Est Cr\'{e}teil, CNRS, ICMPE (UMR 7182), 2 rue Henri Dunant, 94320 Thiais, France}
  \address[krakow]{Institute of Metallurgy and Materials Science of the Polish Academy of Sciences, 30-059 Krakow, Poland}
  
  \begin{abstract}
    Tracer diffusion of all constituting elements is studied at various temperatures in a series of (CoCrFeMn)$_{100-x}$Ni$_x$ alloys with compositions ranging from pure Ni to the equiatomic CoCrFeMnNi high-entropy alloy. At a given homologous temperature, the measured tracer diffusion coefficients change non-monotonically along the transition from pure Ni to the concentrated alloys and finally to the equiatomic CoCrFeMnNi alloy. This is explained by atomistic Monte-Carlo simulations based on a modified embedded-atom potentials, which reveal that local heterogeneities of the atomic configurations around a vacancy cause  correlation effects and induce significant deviations from  predictions of the random alloy model.
  \end{abstract}
  
  \begin{keyword}
    high entropy alloy; diffusion; bulk diffusion; KMC; Random alloy model; Tracer diffusion measurement;
  \end{keyword}
  
\end{frontmatter}


\section{Introduction}

Traditionally, alloys have been developed according to a `base element' paradigm, where one element of the alloy is predominant, e.g. nickel in super alloys or simply iron in steel, and other elements are taken to improve their properties~\cite{Hummel1998}. Since 2004, a new concept, proposed by Yeh~\textit{et al.}.~\cite{Yeh2004a} and Cantor~\textit{et al.}~\cite{Cantor2004}, has gained greater attention, and alloys with multiple principal elements have been investigated. The concentration of each element in these alloys varies between 5 and 35 at.\% but still forms a solid solution. These alloys are commonly known as multi-principal-element alloys or further as high-entropy alloys (HEAs)~\cite{Miracle2017}. The last name goes back to Yeh~\textit{et al.}~\cite{Yeh2004a} because the sheer amount of different elements significantly increases the entropy of mixing $\Delta S_{mix}$. Some of these HEAs have exhibited very promising mechanical properties~\cite{Lilensten2018,Gludovatz2014}, thus initiating a very dynamic field of research. Step by step, properties of HEAs are explored and one of primary importance is diffusion and the activation barriers for kinetic processes.

A concept of 'sluggish' diffusion in HEAs in comparison to conventional alloys was proposed as one of the four `core effects' of HEAs~\cite{Yeh2006}. The first interdiffusion measurements of the CoCrFeMn$_{0.5}$Ni alloys seemed to support this paradigm~\cite{Tsai2013b}. Nowadays, the concept of 'sluggish' diffusion has been questioned~\cite{Vaidya2016,Pickering2016,Vaidya2017,Miracle2017,Divinski2018}. Moreover, the latest diffusion measurements on HEAs have shown that the determination whether diffusion in HEAs is sluggish or not is not straightforward~\cite{Vaidya2018} and highly debated~\cite{Paul2017,Belova2019}. In fact, the type of elements which are involved in the alloy seems to play a more important role than the configurational entropy~\cite{Osetsky2018}. For recent reviews of the available diffusion data in HEAs see, e.g. \cite{Divinski2018, Divinski2020, Dabrowa2020}.

Due to the high configuration complexity of HEAs atomistic computational studies of diffusion in the concentrated solid solutions are rare~\cite{Divinski2018,Choi2018,Mizuno2019} and mostly focus on the equimolar Cantor alloy.
Choi~\textit{et al.}~\cite{Choi2018} sampled the vacancy migration energy for \num{390} vacancy jumps in the equimolar Cantor alloy using their recently developed empirical interatomic potential. They report a broad distribution of migration barriers with a maximum between \SIrange{0.67}{0.87}{\electronvolt} and vacancy formation energies in the range of \SIrange{0.694}{1.207}{\electronvolt}. The calculated hierarchy of migration barriers for different diffusing elements closely matches the experimental diffusivities reported by Tsai~\textit{et al.}~\cite{Tsai2013b}. Therefore, these authors conclude, that the broad energy distribution can lead to chemical environments in the HEA where the vacancy gets trapped, which would in turn reduce the vacancy diffusivity and support claims of sluggish diffusion. 

Vacany migration and formation energies were also studied by Mizuno~\textit{et al.}~\cite{Mizuno2019} based on caclulations within electronic density functional theory (DFT). They calculated the vacancy migration energies for each element in \num{6} different chemical environments using DFT and the previously mentioned interatomic potential and found good agreement between both methods. Contrary to the classical model by Choi~\textit{et al.}~\cite{Choi2018} the average vacancy formation energies calculated by DFT are almost identical for all elements ranging from \num{1.93} to \SI{2.06}{\electronvolt}. However, neither the interatomic potential calculations~\cite{Choi2018} nor the DFT calculations~\cite{Mizuno2019} are in line with the vacancy formation enthalpy in the Cantor alloy measured by positron annihilation spectroscopy of \SI{1.69+-0.13}{\electronvolt}, \SI{1.72+-0.18}{\electronvolt}~\cite{Sugita2020}, or \SI{0.64}{\electronvolt}~\cite{Huang2019}.

In order to obtain a complete understanding of diffusional transport in solids,  statistical  Monte Carlo (MC) simulations, which  can trace many elementary vacancy jumps, are a powerful tool. 
In the context of high entropy alloys, the random alloy model has been used  by Belova~\textit{et al.}~\cite{Belova2000,Belova2001} assuming that species and vacancies are randomly distributed, which implies that average values can be used to represent the actual atom and vacancy jump frequencies in the lattice~\cite{Manning1971}. The results of such simulations can express experimentally measurable
tracer diffusion coefficients which especially allows direct comparison with experiments. However, the random alloy model neglects explicitly the environmental dependence of the jump frequencies and the introduced errors have to be estimated yet.

In the present paper, we have combined tracer diffusion measurements and a novel type of Monte-Carlo simulations,
in order to clarify the `sluggish diffusion' core effect and to understand kinetic processes in HEAs. In doing so, we study diffusion rates of elements in multi-component alloys within the Cantor system for compositions varied from a pure metal to a concentrated solid solution and finally to a HEA, following the approach developed earlier by Laurent-Brocq~\textit{et al.}  to study the thermodynamic stability~\cite{Laurent-Brocq2016} and the solid solution strengthening~\cite{Laurent-Brocq2017,Bracq2019}. Thus, the influence of the concentration of each element on diffusion in HEAs can be determined without having to take into account the influence of the chemical nature of the elements. The present investigation extends the previous experimental study of tracer diffusion in the Ni--CoCrFeMn alloys \cite{Kottke2019} which was conducted at a single temperature of 1373~K.
\begin{sloppypar}
The tracer diffusion of all constituting elements is measured in Co$_{10}$Cr$_{10}$Fe$_{10}$Mn$_{10}$Ni$_{60}$ and Co$_{2}$Cr$_{2}$Fe$_{2}$Mn$_{2}$Ni$_{92}$ alloys in an extended temperature interval. These results are compared to already existing diffusion data for Co$_{20}$Cr$_{20}$Fe$_{20}$Mn$_{20}$Ni$_{20}$ and pure nickel. In order to complete the diffusion databases, we measured Mn diffusion in pure Ni, too. Thus, diffusion in a series of  (CoCrFeMn)$_{100-x}$Ni$_x$ alloys is evaluated ($20\le x \le 100$). All chosen materials were already proven to form single face-centered cubic (FCC) solid solutions~\cite{Laurent-Brocq2016,Bracq2017,Laurent-Brocq2017}. To improve the understanding of the obtained trends in the diffusion constants, atomistic computer simulations are performed, sampling the concentration and chemical environment specific vacancy migration barrier in the Cantor subsystems. These barriers are used as an input for a novel, custom-built kinetic Monte Carlo code. The diffusion constants determined from this code are compared to the experimental results. Moreover, the impact of the migration barrier distribution on the tracer correlation factors is determined comparing the results of our KMC code to the ones determined from the established random alloy model.
\end{sloppypar}

\section{Experimental details}

The Co$_{10}$Cr$_{10}$Fe$_{10}$Mn$_{10}$Ni$_{60}$ and Co$_{2}$Cr$_{2}$Fe$_{2}$Mn$_{2}$Ni$_{92}$ alloys were cast by high frequency electromagnetic induction melting in a water-cooled copper crucible under He atmosphere. Co, Cr, Fe, Mn and Ni metal pieces of at least 99.95~\% purity were melted together and suction-cast as rod-shaped ingots with a diameter of 13~mm. Subsequently, the rods were wrapped in Ta foil, annealed at 1373~K for 13~h under a He atmosphere for chemical homogenization and rapidly quenched afterwards. The chemical composition was examined by energy dispersive X-ray spectroscopy (EDS) at several positions. The determined averaged concentrations are summarized in Appendix, Table~\ref{tab:EDS}.

After homogenization, the Co$_{10}$Cr$_{10}$Fe$_{10}$Mn$_{10}$Ni$_{60}$ and Co$_{2}$Cr$_{2}$Fe$_{2}$Mn$_{2}$Ni$_{92}$ samples were cut by spark erosion and pre-annealed at the diffusion temperatures (wrapped in Ta foil and sealed in quartz ampules under purified Ar atmosphere) to ensure thermal equilibrium conditions. A further set of sample was annealed at 1173~K for X~ray examination. A single face-centred cubic phase at all temperatures was confirmed by X-ray diffraction (XRD) (see Fig.~\ref{fig:XRD} in Appendix).
  
A few microliters of a mixture of $\gamma$-isotopes, i.e. of $^{51}$Cr, $^{54}$Mn, $^{57}$Co and $^{59}$Fe, were dropped on a mirror-like polished surface of the samples. A second set of identical samples was used for diffusion experiments with the $^{63}$Ni $\beta$-isotope. Both sets were diffusion annealed at the selected temperatures for given times (see the experimental conditions listed in Table~\ref{tab:diffcoeff}). Afterwards, the specimens were reduced in diameter to exclude an influence of surface and lateral diffusion. The penetration profiles were determined measuring the relative specific activity of successive sections, which were  parallel-sectioned using a custom-built precise mechanical grinding machine. The thickness of each section was determined from the sample mass difference before and after each grinding step. 

Diffusion on $^{54}$Mn in pure Ni was measured in a similar way.

\section{Details of atomistic calculations}

\subsection{Nudged elastic band calculations}

The nudged elastic band (NEB) method~\cite{Henkelman2000,Henkelman2000a} is used to determine the vacancy migration energy in the different Cantor alloy subsystems. For each (CoCrFeMn)\(_{100-x}\)Ni\(_x\) sample with \(x = 20, 60, 80\) and \(92\) a FCC single crystal containing \num{4000} lattice sites is created. The different species are distributed randomly on these sites. 
The single crystalline samples are first statically relaxed to zero pressure with a maximum force tolerance of \SI{e-8}{\electronvolt\per\angstrom} using conjugate gradient energy minimization. For the following NEB calculations, initial, intermediate, and final states are minimized at constant volume to an energy tolerance of \SI{e-8}{\electronvolt} using the \textsc{fire} algorithm~\cite{Bitzek2006}.
To sample a large number of chemical environments around the vacancy and the migrating atom we now iteratively remove one atom from the sample and calculate the migration barrier for all neighboring atoms jumping in the created vacancy. Once these 12 migration barriers are calculated the removed atom is reinserted and a different atom in the sample is removed. This process is repeated until all lattice sites have hosted the vacancy. Forward and backwards jumps are not calculated twice as they have symmetrical migration energy barriers.

For these calculations \textsc{lammps} is used~\cite{Plimpton1993}. The atomic interactions are described by an empirical second nearest neighbor modified embedded atom interatomic potential parametrized by Choi~\textit{et al.}~\cite{Choi2018}. All samples are created using \textsc{atomsk}~\cite{Hirel2015} and necessary pre- and postprocessing is done in \textsc{ovito}~\cite{Stukowski2010}.
 
\subsection{Kinetic Monte Carlo simulations}

\begin{figure}[tbp]
  \centering
  \includegraphics{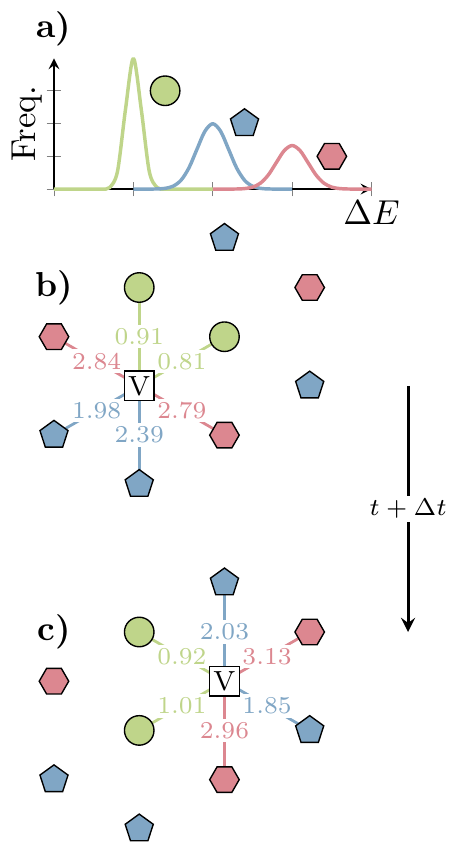}
  \caption{
  Exemplary migration energy barrier distribution in (a) showing three different species with different mean barriers and standard deviation. \(\Delta E\) is the migration barrier and Freq. denotes the frequency for each energy occurring. During each KMC step each neighbor of the vacancy `V' is assigned a migration energy drawn from the distribution (a). A migration event is select based on the rejection free KMC formalism, the event is carried out and the clock is advanced (b $\rightarrow$ c). The energy assignment is repeated for the next KMC step (c). The numbers in (b) and (c) represent the particular values of the migration barriers for possible vacancy jumps.
  }\label{fig:kmc_schematic}
\end{figure}

For this study we developed a novel rejection-free kinetic Monte-Carlo~\cite{Fichthorn1991} code, where migration energies are individually picked from a calculated gaussian distribution. It is assumed that this distribution of migration barriers depends solely on the type of the migrating atom and the overall sample composition (\textit{cf.} Fig.~\ref{fig:Emig}). The barriers used are given in Tab.~\ref{tab:migrationBarrier}. Each KMC step consists of the following operations~\cite{bortz1975}:
\begin{enumerate}
    \item Determine the atomic species in the first nearest neighbor shell of the vacancy.
    \item For each neighbor a migration energy \(\Delta E_i\) is drawn from its respective distribution of migration energies.
    \item Calculate the rate \(\Gamma_i\) for each neighbor exchanging site with the vacancy following
    \begin{equation}
        \Gamma_i = \nu_0 \exp \left( - \frac{\Delta E_i}{k_\mathrm{B} T} \right),
    \end{equation}
    where \(\nu_0 = \SI{e13}{\per\second}\) is the attempt frequency, \(k_\text{B}\) is the Boltmzann constant, \(T\) is the temperature.
    \item The total rate \(\Gamma_\text{tot}\) is equal to
    \begin{equation}
        \Gamma_\text{tot}= \sum_i \Gamma_i.
    \end{equation}
    \item A random number \(u \in (0,1]\) is drawn and an event is selected from the rate catalog so that \(\Gamma_{i-1} < u \Gamma_\text{tot} \leq \Gamma_i\).
    \item The event is applied to the system, \textit{i.e.} the selected atom is exchanged with the vacancy resulting in a diffusive jump.
    \item A second random number \(u'\) is drawn and the time \(t\) is advanced by \(\Delta t\),
    \begin{equation}
    \Delta t = \frac{1}{\Gamma_\text{tot}} \ln \left( 1/u' \right).
    \end{equation}
\end{enumerate}
Figure~\ref{fig:kmc_schematic} shows an example for the migration barrier \(\Delta E\) distribution for three species (a) and a schematic of a KMC step (b-c). In (b) each neighbor of the vacancy `V' gets assigned a migration energy drawn from the distribution. Going from (b) to (c) an event is selected and carried out, the clock is advanced by \(\Delta t\). Now, each neighbor around the vacancy position receives a different energy barrier which is drawn from the energy distribution. These steps are repeated \(10^8\) times and we obtain the trajectories for atoms and the vacancy, as well as the simulated time.

As this modelling approach does not account for the site-specific local chemical environments but only for a statistically correct distribution of all possible environments, the detailed balance criterion is not fulfilled for the individual jumps. Nevertheless, a large number of jumps leads the system towards a global equilibrium.

\begin{table*}
	\renewcommand{\arraystretch}{1.5}
	\footnotesize\centering
	\caption{Temperatures $T$ and times $t$ of the diffusion experiments and the determined diffusion coefficients, $D^*$, for each material and isotope. A typical uncertainty of the determined diffusion coefficients does not exceed $\pm20$\%.
		\label{tab:diffcoeff}}
	\begin{tabular}{c|c|c|c|c|c|c|c}
		\hline
		\multirow{2}{*}{material} & \multirow{2}{*}{$T$ (\si{\kelvin})} & \multirow{2}{*}{$t$ (\si{\second})} & \multicolumn{5}{c}{$D^*$ (\SI{}[10^{-17}]{\meter\squared\per\second})}\\
		\cline{4-8}
		& & & \multicolumn{1}{c|}{$^{51}$Cr} & \multicolumn{1}{c|}{$^{54}$Mn} & \multicolumn{1}{c|}{$^{57}$Co} & \multicolumn{1}{c|}{$^{59}$Fe} & $^{63}$Ni \\ \hline
		\multirow{5}{*}{Co$_{10}$Cr$_{10}$Fe$_{10}$Mn$_{10}$Ni$_{60}$} & 1123 & 1036800 & 1.1 & 2.4 & 1.2 & 1.5 & 0.24 \\
		& 1173 & 933120 & 5.0 & 8.1 & 3.7 & 6.2 & 2.6 \\
		& 1223 & 604200 & 17.1 & 12.4 & 13.8 & 20.7 & 19.1 \\
		& 1273 & 501120 & 19.0 & 65.2 & 24.1 & 38.0 & 10.9 \\ \hline
		\multirow{5}{*}{Co$_{2}$Cr$_{2}$Fe$_{2}$Mn$_{2}$Ni$_{92}$} & 1123 & 1036800 & 1.6 & 3.5 & 1.1 & 2.0 & 1.0 \\
		& 1173 & 933120 & 6.6 & 6.9 & 4.8 & 10.9 & 5.0 \\
		& \multirow{2}{*}{1223} & 604200 & 10.6 & 25.7 &  & 10.7 & \\
		&  & 57600 &  &  & 9.9 &  & 10.5 \\
		& 1273 & 501120 & 30.9 & 33.3 & 20.0 & 67.1 & 49.8 \\ \hline
		\multirow{4}{*}{Ni} & 1123 & 1036800 &  & 2.5 &  &  &  \\
		& 1173 & 933120 &  & 7.0 &  &  &  \\
		& 1223 & 604200 &  & 22.2 &  &  &  \\
		& 1273 & 503712 &  & 48.4 &  &  &  \\ \hline
	\end{tabular}
\end{table*}

All following KMC results are obtained from a sample containing \num{32000} lattice sites with a single vacancy. These lattice sites are randomly filled with atoms in the desired concentration. Periodic boundary conditions are employed to approximate an infinitely large sample.

Using a single valued energy barrier distribution in this KMC code gives identical properties as the conventional random alloy model. We used this setup as a benchmark and compared the correlation factors obtained from the newly developed code after \(10^6\) and \(10^8\) simulation steps to the \(f\) values calculated from a conventional random alloy model after \(10^{12}\) steps. The results of this comparsion are shown in Fig.~\ref{fig:Appendix_f} in the Appendix

The tracer correlation factor, \(f\), influences the tracer diffusion coefficient, \(D^*\),
\begin{align}
  D^* &= f D,  \label{eq:D_sim_1}\\
  D & = c_\text{Vac} \nu_0 \lambda^2 \exp \left( \frac{-\Delta E_\text{Mig}}{k_\text{B}, T}\right) \label{eq:D_sim_2}
\end{align}
where \(D\) is the bulk diffusivity and $c_{\rm Vac}$ is the vacancy concentration \cite{Mehrer2007}. The correlation factor can be obtained from KMC simulations by counting the number of jumps \(n_i\) and the displacement \(R_i\) of each atom \(i\). \(f\) is given then as,
\begin{equation}
  f = \frac{\sum_i R_i^2}{\sum_i n_i \lambda^2},
\end{equation}
where \(\lambda\) is the jump length of a diffusional jump~\cite{Murch1984}.

\section{Results and discussion}

\subsection{Volume diffusion}

\begin{figure}[ht]
	\begin{center}
		\includegraphics{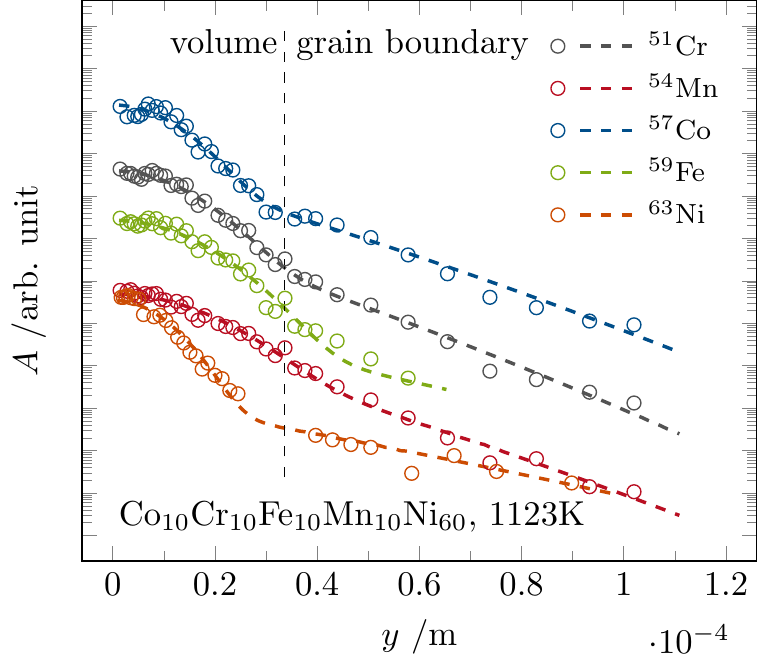}
	\end{center}
	\caption{Measured tracer penetration profiles of $^{51}$Cr (black), $^{54}$Mn (red), $^{57}$Co (blue), $^{59}$Fe (olive) and $^{63}$Ni (orange) in Co$_{10}$Cr$_{10}$Fe$_{10}$Mn$_{10}$Ni$_{60}$ after a heat treatment at 1123 K for 10.6 days. Each circle symbolizes one data point at a specific depth $y$. The dashed lines correspond to the Suzuoka solutions~\cite{Suzuoka1964} of the grain boundary diffusion problem accounting for both, volume and short-circuit diffusion. The profiles are shifted along the ordinate axis for a better visualization.
		\label{fig:profile}}
\end{figure}

In this study, we focus on volume diffusion even though short-circuit diffusion paths were occasionally also observed. 
The experimental conditions (temperatures $T$ and times $t$) of the diffusion experiments are summarized in Table~\ref{tab:diffcoeff}.
Examples of the measured concentration profiles are shown in Fig.~\ref{fig:profile}, where the penetration profiles of all constituting elements in Co$_{10}$Cr$_{10}$Fe$_{10}$Mn$_{10}$Ni$_{60}$ at \SI{1123}{\kelvin} are presented. Two contributions, representing volume and grain boundary diffusion at near-surface and deeper depths, respectively, can clearly be distinguished in this particular case.

The volume diffusion coefficients can be extracted by plotting the logarithm of the relative specific activity, \(A\), of the tracer against the diffusion depth squared, \(y^{2}\). In this representation, a linear decrease of the tracer concentration is observable up to a certain depth, depending on the diffusion temperature and the annealing time. For the analysis, the instantaneous source solution of the diffusion problem is used~\cite{Paul2014},

\begin{figure*}
	\includegraphics[width=0.95\linewidth]{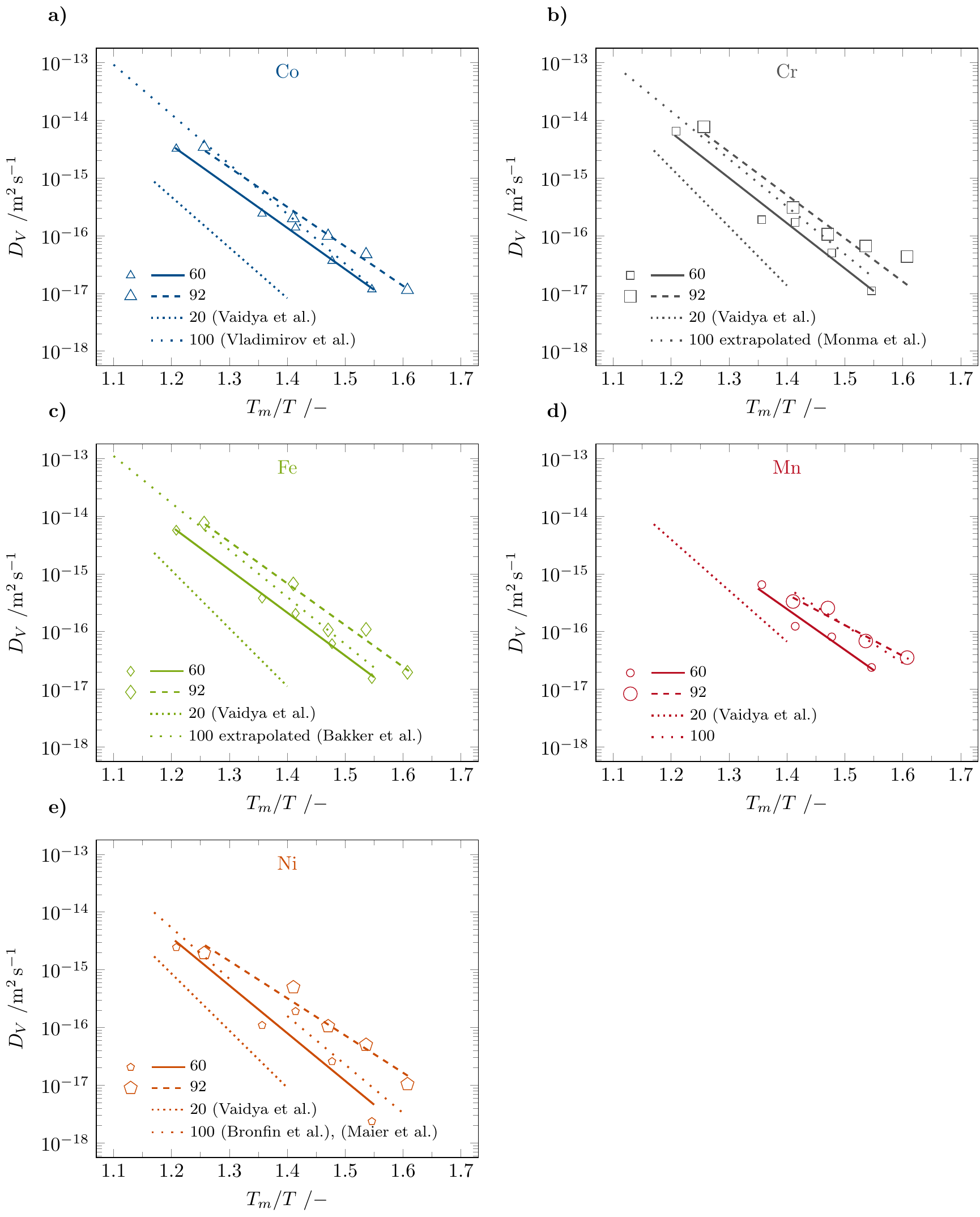}
	\caption{The measured volume diffusion coefficients against the inverse homologous temperature, $T_m/T$, where \(T\) is the diffusion temperature and \(T_m\) is the melting temperature, for $^{57}$Co (a), $^{51}$Cr (b), $^{59}$Fe (c), $^{54}$Mn (d), and $^{63}$Ni (e) diffusion in Co$_{10}$Cr$_{10}$Fe$_{10}$Mn$_{10}$Ni$_{60}$~(60) and Co$_{2}$Cr$_{2}$Fe$_{2}$Mn$_{2}$Ni$_{92}$~(92) alloys in comparison to pure Ni~(100)~\cite{Vladimirov1978,Monma1964,Bakker1971,Bronfin1975,Maier1976} and the Co$_{20}$Cr$_{20}$Fe$_{20}$Mn$_{20}$Ni$_{20}$ HEA~(20)~\cite{Vaidya2016,Vaidya2018}. Note that the diffusion coefficients at \SI{1373}{\kelvin} were measured in our previous work~\cite{Kottke2019} and Mn diffusion in pure Ni was measured in the present work, too. The diffusion coefficients in the two alloys are distinguished by the symbol size. 
		\label{fig:arrhenius}}
\end{figure*}

\begin{equation}
A(y,t) = \frac{M}{\sqrt{4\pi D^*t}}\exp\left( -\frac{y^2}{4D^*t}\right).
\label{eq:volumeinstantaneous}
\end{equation}

Here, $M$ is the initial amount of tracer applied onto the surface and $t$ is the diffusion annealing time. Consequently, the volume diffusion coefficients, $D^*_{i}$, of each tracer $i$ can directly be determined by fitting the relevant parts of the penetration profiles according to Eq.~(\ref{eq:volumeinstantaneous}). Such contributions are seen, e.g., in Fig.~\ref{fig:profile} till the depths of about \SI{20}{\micro\meter}.

In some cases of tracer deposition, an increased amount of the remnant tracer was found on the sample surface. Hence, the constant source solution had to be applied for the volume part of the profile,

\begin{equation}
A = A_0 \erfc\left(\frac{y}{\sqrt{4 D_{\mathrm{V}}t}}\right).
\label{eq:volumeconstant}
\end{equation}

\subsection{Grain boundary diffusion}

The grain boundary diffusion contribution has to be evaluated depending on the specific kinetic regime as it was introduced by Harrison~\cite{Harrison1961}. In the present study, the diffusion measurements were performed in the B-type kinetic regime and we used the exact solution of the Fisher model~\cite{Fisher1951} for an instantaneous source, which was proposed by Suzuoka~\cite{Suzuoka1964}. 

An analysis of the experimental diffusion profiles is exemplified in Fig.~\ref*{fig:profile} where contributions of volume and grain boundary diffusion could directly be distinguished. 

The careful analysis of the grain boundary diffusion contributions allowed reliably to obtain the volume diffusion coefficients which are of prime importance in the present paper. To this end, the penetration profiles were fitted accounting for both, volume [Eqs.~(\ref{eq:volumeinstantaneous}) and (\ref{eq:volumeconstant})] and grain boundary [Suzuoka solution \cite{Suzuoka1964}] diffusion fluxes. The determined volume diffusion coefficients are given in Table~\ref{tab:diffcoeff}.

\subsection{Temperature dependence}

The measured diffusion coefficients of all elements in Co$_{10}$Cr$_{10}$Fe$_{10}$Mn$_{10}$Ni$_{60}$ and Co$_{2}$Cr$_{2}$Fe$_{2}$Mn$_{2}$Ni$_{92}$ are plotted against the inverse homologous temperature \(T_m/T\) where \(T\) is the diffusion temperature and \(T_m\) is the melting temperature. in Fig.~\ref{fig:arrhenius}. The melting temperatures of Co$_{10}$Cr$_{10}$Fe$_{10}$Mn$_{10}$Ni$_{60}$ is \SI{1659}{\kelvin} and \SI{1743}{\kelvin} for Co$_{2}$Cr$_{2}$Fe$_{2}$Mn$_{2}$Ni$_{92}$~\cite{Kottke2019}.
Typically, the change of the tracer diffusion coefficients over a specific temperature range can be described by an Arrhenius dependence~\cite{Mehrer1990},

\begin{equation}
D^* = D^*_{0} \exp \left(-\frac{Q}{RT}\right),
\label{eq:arrhenius}
\end{equation}

\noindent	with the pre-exponential factor~$D^*_{0}$ and the activation enthalpy~$Q$. The determined activation enthalpies for bulk diffusion of all constituting elements in the alloys under consideration are summarized in Table~\ref{tab:diffusion}.

As one can see, Mn is found to be the fastest element which features the lowest activation enthalpy for bulk diffusion in these alloys. The Ni atoms have shown the highest activation enthalpy, being larger than the activation enthalpy for Mn bulk diffusion by a factor of 1.5. Further, our experiments indicate that diffusion of Fe and Cr can be treated fairly similarly. 

In Fig.~\ref{fig:arrhenius}, the presently measured tracer diffusion coefficients are compared to those in pure nickel and in the Cantor alloy (Co$_{20}$Cr$_{20}$Fe$_{20}$Mn$_{20}$Ni$_{20}$). Note that Mn diffusion in pure Ni was measured in the present investigation, too. The experiments reveal that the diffusion rates in the investigated alloys can be ordered as follows: \\
Co$_{2}$Cr$_{2}$Fe$_{2}$Mn$_{2}$Ni$_{92}$ $>$ Ni $>$ Co$_{10}$Cr$_{10}$Fe$_{10}$Mn$_{10}$Ni$_{60}$ $>$\\
Co$_{20}$Cr$_{20}$Fe$_{20}$Mn$_{20}$Ni$_{20}$.

Figure~\ref{fig:arrhenius} suggests that, if analysed at a given homologous temperature, alloying of pure Ni by 8\% of an equiatomic mixture of CoCrFeMn, i.e. by 2\% of each element, enhances the diffusion rates of all elements, excluding Mn. The Mn diffusion rates are almost unaffected by this alloying, though a minor tendency towards retardation at lower $T/T_m$ values ($T/T_m < 0.7$) and an acceleration at higher temperatures, $T/T_m > 0.7$, might be indicated.

Alloying by \SI{40}{\percent} of an equiatomic mixture of CoCrFeMn decreases the diffusion coefficients to the level typical for element diffusion in pure Ni. Finally, somewhat sluggish diffusion in CoCrFeMnNi HEA might be stated, especially for Fe. Again, Mn is an exception from this 'rule'. When extrapolated to the melting point, $T/T_m \approx 1$, the diffusion coefficient of Mn in CoCrFeMnNi HEA becomes even higher than that in pure Ni or at least very similar values are observed.

\begin{table*}[t]
  \renewcommand{\arraystretch}{1.3}
  \scriptsize
  \centering
  \caption{The vacancy migration energies, $E_{\mathrm{vac}_m}$, determined by atomistic simulations and the measured diffusion activation enthalpies, $Q$, for Co$_{10}$Cr$_{10}$Fe$_{10}$Mn$_{10}$Ni$_{60}$ and Co$_{2}$Cr$_{2}$Fe$_{2}$Mn$_{2}$Ni$_{92}$. All data is given in \si{\kilo\joule\per\mole}. For comparison, the data for pure Ni and the CoCrFeMnNi HEA are presented, too. Note that Mn diffusion in pure Ni was measured in the present work.}\label{tab:diffusion}
  \begin{tabular}{l|c|c|c|c|c|c|c|c}
          \hline
          \multirow{2}{*}{Tracer} &
          \multicolumn{2}{c|}{Co$_{20}$Cr$_{20}$Fe$_{20}$Mn$_{20}$Ni$_{20}$} & \multicolumn{2}{c|}{Co$_{10}$Cr$_{10}$Fe$_{10}$Mn$_{10}$Ni$_{60}$} &
          \multicolumn{2}{c|}{Co$_{2}$Cr$_{2}$Fe$_{2}$Mn$_{2}$Ni$_{92}$} & \multicolumn{2}{c}{Ni} \\ \cline{2-9}
          & $E_{\mathrm{vac}_m}$ & $Q$ & $E_{\mathrm{vac}_m}$ & $Q$  &  $E_{\mathrm{vac}_m}$ & $Q$ & $E_{\mathrm{vac}_m}$ & $Q$ \\ \hline
          $^{51}$Cr & 54.3  & $313 \pm 13$~\cite{Vaidya2018} & 89.5  & $249 \pm 24$ & 114.7 & $247 \pm 19$ & 158.8  & $272.6$~\cite{Monma1964} \\
          $^{54}$Mn & 53.4  & $272 \pm 13$~\cite{Vaidya2018} & 89.7  & $224 \pm 38$ & 116.0 & $175 \pm 28$ & 140.1 & $219 \pm 8$ \\ 
          $^{57}$Co & 62.9 & $270 \pm 22$~\cite{Vaidya2018} & 101.0 & $227 \pm 9$ & 124.6 & $226 \pm 13$ & 160.3 &  $285.1$~\cite{Vladimirov1978} \\
          $^{59}$Fe &  56.1 & $309 \pm 11$~\cite{Vaidya2018} & 89.1 & $237 \pm 10$ & 113.1 & $238 \pm 25$ & 138.1 &  $269.4$~\cite{Bakker1971} \\
          \multirow{2}{*}{$^{63}$Ni} & \multirow{2}{*}{62.2} & \multirow{2}{*}{$304 \pm 9$~\cite{Vaidya2016}} & \multirow{2}{*}{96.7} & \multirow{2}{*}{$262 \pm 47$} & \multirow{2}{*}{118.4} & \multirow{2}{*}{$213 \pm 24$} & \multirow{2}{*}{142.7} & $292.6$~\cite{Bronfin1975} \\
          & & & & & & & & $280.8$~\cite{Maier1976} \\
          \hline
        \end{tabular}
      \end{table*}

\begin{table*}[ht]
	\centering
	\caption{Mean, \(\mu\), and standard deviation, \(\sigma\), of the distribution of the  activation energies obtained from the NEB calculations shown in Fig.~\ref{fig:Emig}. \(x_{\text{Ni}}\) is given in \si{\atpercent}, \(\mu\) and \(\sigma\) are given in \si{\electronvolt}/{\rm at}. \(x_{\text{Ni}}=100~\si{\atpercent}\) corresponds to the case where a single solute is added to the sample. This data is used as input for the KMC simulations.
	}\label{tab:migrationBarrier}
	\begin{tabular}{l | r r | r r | r r | r r | r r}
		\hline
		\(x_{\text{Ni}}\) & \multicolumn{2}{c|}{Co} & \multicolumn{2}{c|}{Cr} & \multicolumn{2}{c|}{Fe} & \multicolumn{2}{c|}{Mn} & \multicolumn{2}{c}{Ni} \\
		& \(\mu\) & \(\sigma\) & \(\mu\) & \(\sigma\) & \(\mu\) & \(\sigma\) & \(\mu\) & \(\sigma\) & \(\mu\) & \(\sigma\)\\
		\hline
		\num{20} & \num{1.045} & \num{0.224}& \num{0.869}& \num{0.295}& \num{0.690}& \num{0.164}& \num{0.493} & \num{0.168} & \num{0.985} & \num{0.208} \\
		\num{60} & \num{1.310} & \num{0.172} & \num{1.261} & \num{0.267} & \num{0.862} & \num{0.163} & \num{0.707} & \num{0.135} & \num{1.216} & \num{0.166} \\
		\num{80} & \num{1.445} & \num{0.122} & \num{1.465} & \num{0.192} & \num{0.975} & \num{0.140} & \num{0.775} & \num{0.111} & \num{1.339} & \num{0.115}\\
		\num{92} & \num{1.570} & \num{0.079} & \num{1.563} & \num{0.1444} & \num{1.048} & \num{0.083} & \num{0.820} & \num{0.086} & \num{1.421} & \num{0.085}\\
		\num{100} & \num{1.651} & & \num{1.625} & & \num{1.073} & & \num{0.850} && \num{1.471}\\
		\hline
	\end{tabular}
\end{table*}

\subsection{Atomic migration energy barriers}

The vacancy migration energy barriers, \(\Delta E_\text{Mig}\), in a HEA are expected to strongly depend on the chemical environment of the vacancy. All neighboring atoms can jump into the vacant site and given the chemical complexity of the alloy the activation energy for all theses jumps will be different. The determination of the migration energy distribution requires sampling of a great number of chemical environments. 

\begin{figure*}[tb]
	\begin{center}
  \includegraphics[width=0.9\linewidth]{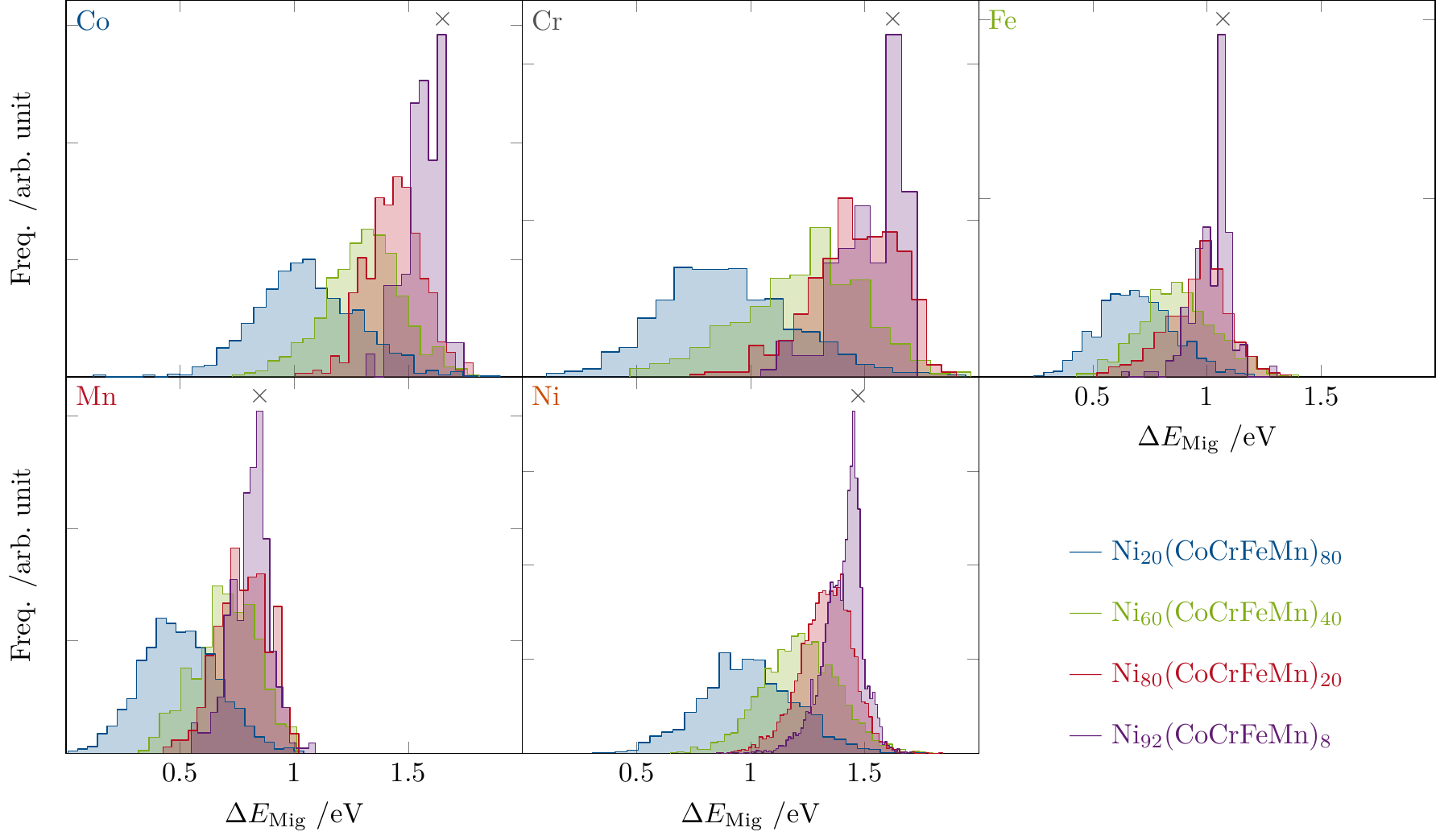}  
	\end{center}
    \caption{Distribution of the vacancy migration energy barriers \(\Delta E_\text{Mig}\) in different Ni\(_x\)(CoCrFeMn)\(_{100-x}\) alloys. Freq. is the frequency of the different \(\Delta E_\text{Mig}\). The label in the top right denotes the species of the migrating atom. The \(\times\) denotes the diffusion activation energy for the migrating element in a pure FCC Ni matrix (corresponding to the dilute limit). The total data set includes about \num{29000} vacancy jumps.
    }\label{fig:Emig}
\end{figure*}

Using the  interatomic potentials of Choi~\textit{et al.}~\cite{Choi2018}, we use the nudged-elastic band method to determine the vacancy migration barrier for different atoms exchanging position with the vacancy. By randomly sampling different chemical environments we obtain the distributions of \(\Delta E_\text{Mig}\)  shown in Fig.~\ref{fig:Emig}. Each plot shows the distribution (number fraction) of the concentration-dependent migration barriers for one migrating species. 
It can be seen that Mn has the lowest mean migration energy barrier combined with a rather narrow energy distribution. Cr has a slightly higher mean migration barrier but shows a significantly broader distribution. Co and Ni both are predicted to have the highest vacancy migration energies. Moreover, as the Ni concentration increases, i.e. with a transition to more dilute Ni-based alloys, the means of all distributions shift to higher energies. Tab.~\ref{tab:migrationBarrier} summarizes the mean of the peak positions and their width, where \(\mu\) denotes the mean of the fitted Gaussian and \(\sigma\) its standard deviation. 

It can also be seen that as the Ni concentration increases, the probability of having a Ni-rich environment around a vacancy also increases leading to a reduction in peak width for all samples. This is also accompanied by a splitting of this major peak. This corresponds to binning of the Ni-rich environments and all other environments around the vacancy.
The dilute-limit migration energy barrier, which corresponds to a single solute atom in a pure Ni matrix, is highlighted by a \(\times\) symbol. It can be seen that this energy corresponds to the main peak in the \(x=92\) sample indicating a high probability for a pure Ni neighborhood.

Comparing our findings with the vacancy migration barriers reported by Choi~\textit{et al.}~\cite{Choi2018} we find the same migration energy distribution if we superimpose the different elemental histograms given in Fig~\ref{fig:Emig}. This results is expected given that both results are based on the same interatomic potential. Nevertheless, the significantly larger number of migration barriers sampled for this study (\num{5604} in the present study against the \num{390} of Choi~\textit{et al.}~\cite{Choi2018} for the equimolar Cantor alloy) allows for a species resolved migration barrier distribution instead of one global migration energy distribution. 

Mizuno~\textit{et al.}~\cite{Mizuno2019} studied the migration barrier using small special quasi-random structures (SQS) using DFT. These SQS are equimolar and designed to mimic a random chemical environment on a small scale. We find that many local chemical environments are not perfectly equimolar or perfectly random given the large number of combinatoric degrees of freedom on the lattice. These fluctuations cannot be captured using small SQS.

\subsection{Tracer correlation factors from KMC simulations}

\begin{figure}[tbp]
\includegraphics{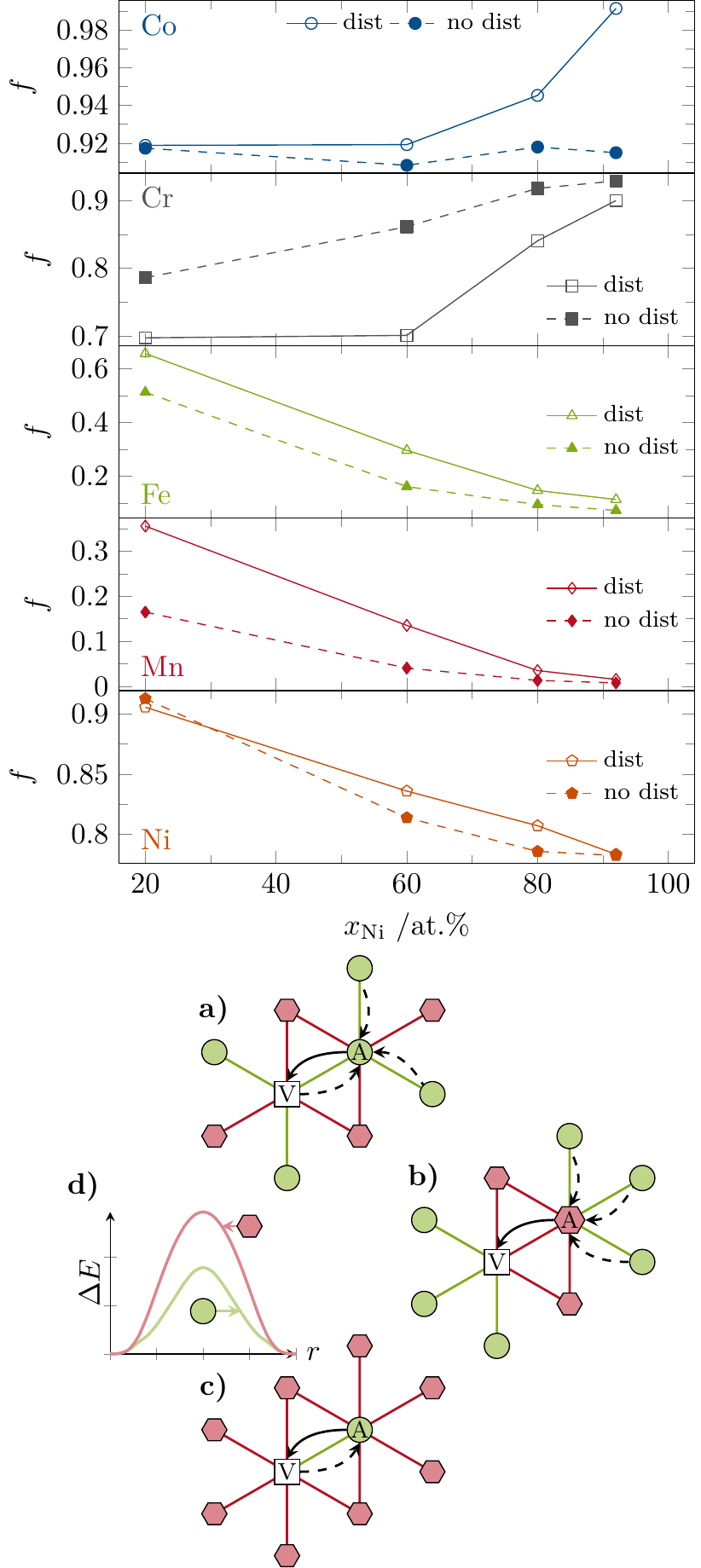}%
 \caption{Correlation factors obtained from the KMC simulation after \(10^8\) steps. Solid lines represent KMC results where the migration energy barriers are taken with a finite width (Tab.~\ref{tab:migrationBarrier}). Dashed lines show \(f\) with a constant, unimodal activation barrier activation barrier, which corresponds to the conventional random alloy model. This data corresponds to the random alloy model. In all cases \(f\) was sampled at \SI{1123}{\kelvin}. (a-c) show examples of different chemical environments around the vacancy. Red bonds correspond to a high migration barrier jump, while green bonds represent low migration barrier jumps. The first jump is indicated by a solid arrow, while most probable successive jumps are indicated by dashed lines. (d) schematically shows the low and high migration energy \(\Delta E\) over the reaction coordinate \(r\). Fig.~\ref{fig:fCantor_dist_nodist_appendix} shows the same \(f\) data in a singular plot for easier comparison of the different samples.
 }\label{fig:fCantor_dist_nodist}
\end{figure}

The solid lines in Fig~\ref{fig:fCantor_dist_nodist} show the tracer correlation factor \(f\) calculated from the vacancy migration energy barrier distributions given in Fig.~\ref{fig:Emig}.
The data shows that the correlation factor for the faster diffusing elements (Mn and Fe) with the lower migration energy barrier is reduced compared to the other elements in the alloy. Moreover, the concentration dependence of \(f\) shows that \(f\) decreases further as the Ni concentration increases. As the Ni concentration approaches 1, the Ni correlation factor tends towards the theoretical \(f\) value for low defect concentrations of 0.78146~\cite{Murch1982}.

Both observations can be explained by looking at the energy landscape around the vacancy. Three different cases are shown schematically in Fig.~\ref{fig:fCantor_dist_nodist} (a-c). Here green bonds correspond to low migration energy jumps, while red bonds indicate high energy jumps (migration energies are schematically shown in (d)). The initial vacancy position is marked by a `V'. The first jump is always `A' and `V' exchanging sites  (solid arrow). In (a) and (c) this would be a low energy barrier jump, while in (b) this requires crossing a high energy barrier. Successive jumps with a low energy barrier and therefore a high probability are indicated by dashed arrows.
From this simplified picture we can now learn that in (a) an atom with a low migration barrier has a high probability of jumping forwards and backwards leading to two diffusive jumps without net mass transport therefore leading to a more correlated tracer diffusivity. From (b) we can see that this successive forwards-backwards jump is less likely for atoms having a high migration energy barrier. These atoms are more likely to jump once and have the vacancy progress further in the crystal instead of jumping backwards leading to a correlation factor closer to one. (c) shows a similar situation in a system where the high migration barrier atoms make up a larger fraction (similar to the \(x_\text{Ni}=92\) sample). Once an atom with a low activation barrier is found by the vacancy the probability of it jumping back and forth around the same position becomes highly probable leading to a highly correlated diffusion trajectory for the low migration barrier species.

\subsection{Diffusion constants from KMC simulations}
The KMC simulations give us the diffusion trajectories of all atoms and the vacancy contained in the sample. Using this data the diffusion coefficient \(D_i\) can be calculated from the mean squared displacement \(R_i\) for each species \(i\) given the relation,
\begin{equation}
  \langle R_i^2 \rangle = \frac{1}{6} D_i t,
\end{equation}
where \(t\) is the total time. In practice \(D_i\) is obtained from a linear fit of \(\langle R_i^2 \rangle\) over \(t\)~\cite{Frenkel2002}. 

\begin{figure*}[tbp]
  \includegraphics{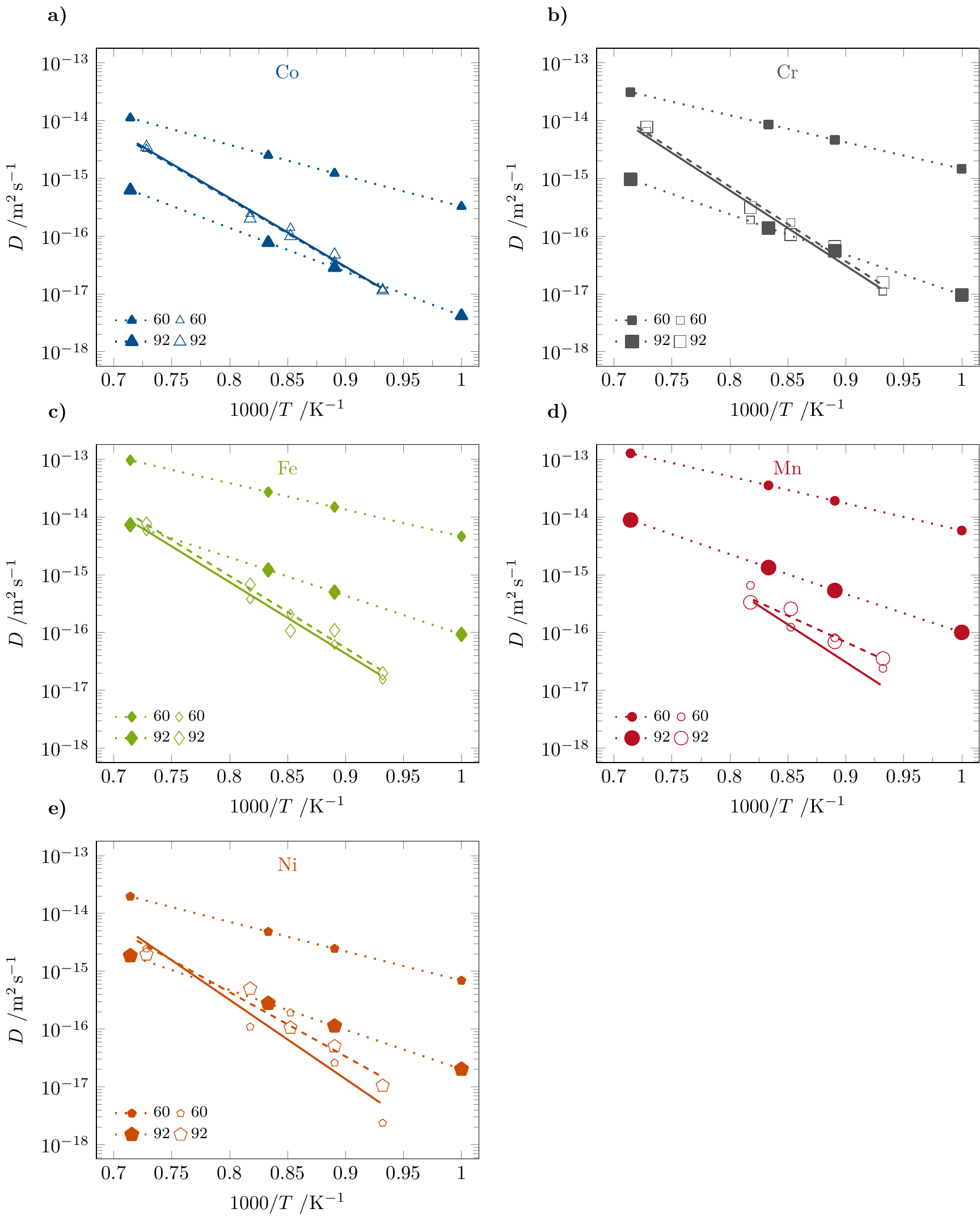}  
    \caption{Arrhenius plots for element-dependent diffusion coefficients (a) -- (e) in Co$_{10}$Cr$_{10}$Fe$_{10}$Mn$_{10}$Ni$_{60}$ (small symbols) and Co$_{2}$Cr$_{2}$Fe$_{2}$Mn$_{2}$Ni$_{92}$ (large symbols). The simulation data (filled symbols and dotted lines) are presented for a constant vacancy concentration of \num{3.125e-5}. The experimentally measured diffusion coefficients are given by open symbols. The solid and dashed lines represent the corresponding Arrhenius fits.
    }\label{fig:DTE}
\end{figure*}

Fig.~\ref{fig:DTE} shows \(D_i\) on an Arrhenius scale for the different elements and Ni concentrations in the sample. We can see that for all samples and all elements \(D_i\) decreases as the Ni concentration in the sample increases. Moreover, the diffusion is always higher in the alloy compared to the dilute limit of a single solute in a pure Ni matrix (\(x_\text{Ni}=100\)).

Further examination of \(D_i\) reveals that it follows the trend one would expect from the activation energy barriers presented in Table~\ref{tab:migrationBarrier} and transition state theory. A high migration energy corresponds to a lower diffusivity while a lower barrier leads to a higher \(D_i\). There are two main factors explaining this relationship. First, as shown in Fig.~\ref{fig:fCantor_dist_nodist}, while there is a contribution of the correlation factor \(f\) to the diffusivity in these alloys, it changes only by about one order of magnitude, which is not sufficient to overcome the differences in \(D_i\) stemming from the significant difference in \(\Delta E_i\) between elements and concentrations. Second, the data shown in Fig.~\ref{fig:DTE} does not account for concentration or temperature dependent differences in the vacancy concentration which would act as a scaling factor on the individual \(D\) values (Eq.~\ref{eq:D_sim_2}). The vacancy formation energies cannot be accounted for as there is no applicable theory on how to average the vacancy formation energies for each species and how to include the concentration dependant configurational entropy of the vacancy.

\subsection{Comparison between experiment and simulation}

To avoid the uncertainties related to the vacancy concentration and the attempt frequencies \(\nu_0\), which have both been defined as the given constants for the KMC simulations, we normalized the element-specific diffusion coefficients, \(D_i\), by the diffusion coefficient of Fe, \(D_\text{Fe}\). From Eq.~(\ref{eq:D_sim_1}) and (\ref{eq:D_sim_2}) it can be seen that \(D_i / D_\text{Fe}\) is obviously independent of \(c_\text{Vac}\) and \(\nu_0\) as they are constant for one sample at a given temperature (neglecting simultaneously the variation of the element-specific migration entropies). The specific choice of Fe is somewhat arbitrary and it is motivated by a lowest scatter of the experimental points on the corresponding Arrhenius plots for all alloys under consideration. The comparison of experimental tracer diffusion coefficients with those obtained from the KMC simulations reveals a good qualitative agreement and the differences are caused probably by the interatomic potentials used in the present work.

\begin{figure}[tbp]
  \includegraphics{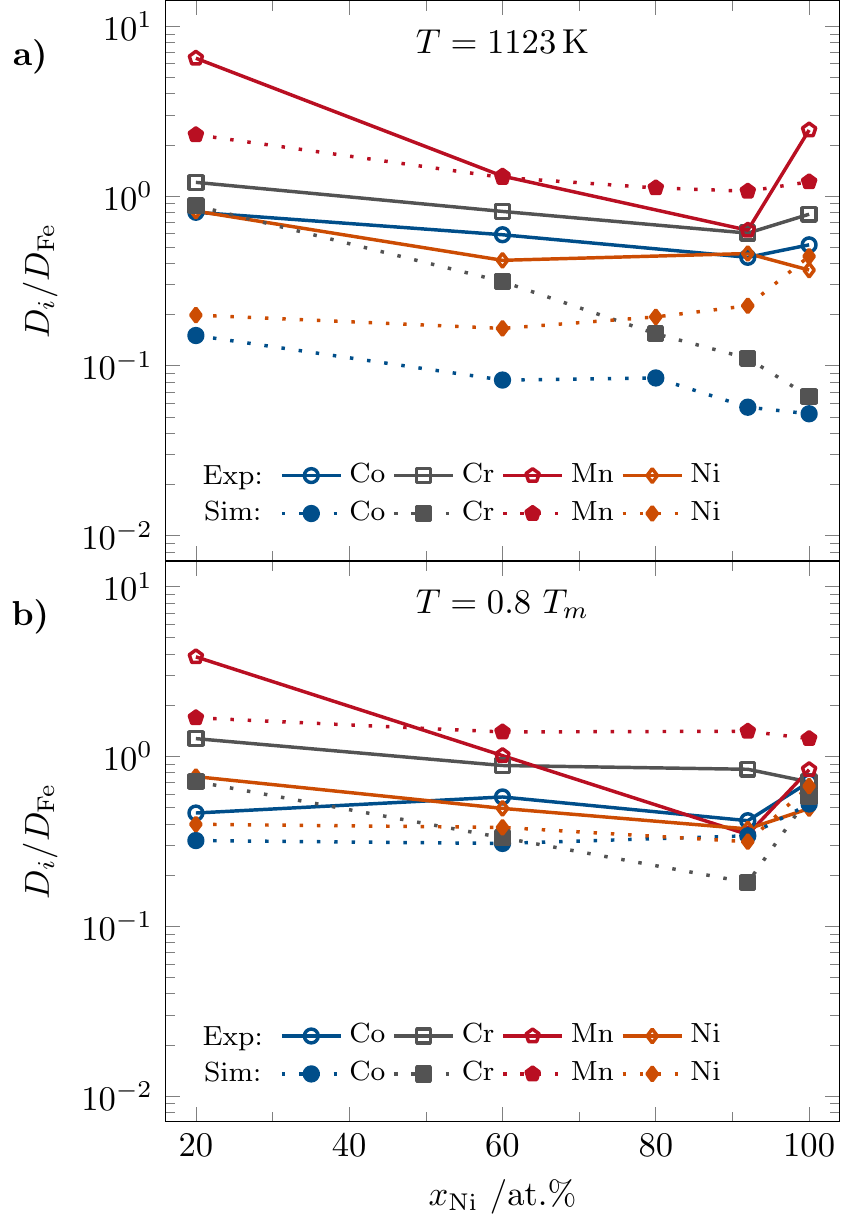}
  \caption{Comparison of element-specific diffusion coefficients normalized by the corresponding diffusion coefficients of Fe, $D_i/D_{\rm Fe}$, as predicted by simulation (filled symbols and dashed lines) and determined experimentally (open symbols and solid lines). In a), the measured and calculated data are compared at \SI{1123}{\kelvin}, whereas in b), the comparison is done at $T=0.8T_m$.
  }\label{fig:DxDFe_1123}
 \end{figure}

Figure~\ref{fig:DxDFe_1123}~a shows \(D_i / D_\text{Fe}\) as a function of the Ni concentration at a constant absolute temperature of \SI{1123}{\kelvin}. A general increasing trend of the normalized diffusion coefficients from pure Ni to the Cantor alloy which is observed for all solutes can be reproduced by our simulations, too. Qualitatively, the experimental and simulation data agree especially well for Mn and Ni, while for Cr a significantly stronger increase of the normalized diffusion coefficient, $D_{\rm Cr}/D_{\rm Fe}$, is predicted, from less than 0.1 (pure Ni) to about unity (HEA). Experimentally, these values are equal to almost 1 for all compositions. The normalized Ni diffusion coefficient, $D_{\rm Ni}/D_{\rm Fe}$, decreases first from about 0.7 (pure Ni) to 0.6 (the concentrated alloy) and than increases to about 0.8 (the Cantor alloy).

When analysed at a constant homologous temperature of $0.8~T_m$, Figure~\ref{fig:DxDFe_1123}~b, the normalized diffusion coefficients group within a relatively narrow interval $0.2 < D_i/D_{\rm Fe} < 2$. The atomistic simulations reproduce well the concentration trends observed for the diffusion of Ni and Co atoms, even quantitatively and for other elements qualitatively correct tendencies are reproduced. 

Mn is probably the most important exception. While atomistic calculations predict $D_{\rm Mn}/D_{\rm Fe} \ge 1$ for all compositions at $T=0.8T_m$, a significant decrease of the normalized diffusion coefficient is seen in concentrated alloys that is reversed in CoCrFeMnNi HEA.

\section{Conclusions}
In the present work, volume diffusion of all constituting elements is measured in Co$_{10}$Cr$_{10}$Fe$_{10}$Mn$_{10}$Ni$_{60}$ and Co$_{2}$Cr$_{2}$Fe$_{2}$Mn$_{2}$Ni$_{92}$ at temperatures from 1123~K to 1373~K. The two alloys were shown to form a single-phase FCC solid solution at all measured temperatures. For completeness, Mn diffusion in pure Ni is measured, too, to provide a whole data set for reliable evaluation of the element-specific diffusion coefficients along the Ni--CoCrFeMnNi cut of the multi-component phase diagram.  Simultaneously, the vacancy formation and migration enthalpies in these alloys are examined using atomistic simulations with empirical interatomic potentials.

Reconsidering the measurements and the simulations, the following conclusions are reached:
\begin{itemize}
    \item The measured activation enthalpies are varying between 175~$\si{\kilo\joule\per\mole}$ and 247~$\si{\kilo\joule\per\mole}$ for\\ 
    Co$_{2}$Cr$_{2}$Fe$_{2}$Mn$_{2}$Ni$_{92}$ and from 224~$\si{\kilo\joule\per\mole}$ to \\
    262~$\si{\kilo\joule\per\mole}$ for Co$_{10}$Cr$_{10}$Fe$_{10}$Mn$_{10}$Ni$_{60}$.
    \item On the inverse homologous temperature scale, the data suggest that bulk diffusion is enhanced by small additions of a solute mixture due to a higher vacancy concentration. Further addition of the equiatomic CoCrFeMn mixture (10~at.\% each) leads to a retardation of the diffusion rates in comparison to solute diffusion in pure Ni. In general, this cannot be taken as a clear evidence for `sluggish' diffusion, since the effect depends strongly on the temperature and especially on the solute.
    \item We employed high-throughput sampling of more than \(29000\) vacancy migration barriers to determine the species and concentration dependent migration barrier distributions. Here, we find that Mn has the lowest energy barrier, followed by Cr or Fe, depending on the sample composition. Co and Ni show the highest barriers. 
    \item We find that the mean of the energy barrier distributions is most strongly affected by the species of the diffusing atom, while the width correlates with the sample composition and decreases as the sample shifts from HEA to dilute solid solution.
    \item The tracer correlation factors for each element in the HEA obtained from our KMC simulations correlate with the respective mean of the migration barrier distribution. Fast diffusing elements with low barriers show highly correlated jumps, especially as the concentration of high vacancy migration barrier elements in the alloy increases.
    \item The finite width of the vacancy migration barrier modifies the tracer correlation factors compared to the conventional random alloy model. Species with low migration barriers of broad migration barrier distributions are impacted most strongly.
    \item The comparison between measurements and simulation indicates that our model provides a reasonable description. The absolute diffusion coefficients can not be reproduced so far, since there is no no applicable theory on how to average the vacancy formation energies for each species and how to further include the concentration dependant configurational entropy of the vacancy. Neglecting the vacancy concentration by normalizing the diffusion coefficients on one diffusor (Fe), all observable trends of the measurements could be replicated.
    \item The element-specific correlation factors are strongly affected by the heterogeneities of local chemical environments of a diffusing vacancy causing strong deviations from predictions of a mean-field random alloy model.
\end{itemize}

\section{Acknowledgments}

Funding by Deutsche Forschungsgemeinschaft (DFG) via SPP2006, projects WI 1899/32-1 and STU 611/2-1, is gratefully acknowledged.
Calculations for this research were conducted on the Lichtenberg high performance computer of the TU Darmstadt.

\appendix
\section{Appendix}

\begin{table*}
\renewcommand{\arraystretch}{1.5}
\footnotesize\centering
\caption{Element concentrations (in at.\%) as determined by EDS analyses of the homogenized rods which are used radiotracer diffusion experiments.
\label{tab:EDS}}
\begin{tabular}{|c|c|c|c|c|c|}
  \hline 
  Sample & Co & Cr & Fe & Mn & Ni \\ 
  \hline 
  Co$_{10}$Cr$_{10}$Fe$_{10}$Mn$_{10}$Ni$_{60}$ & $10.13 \pm 0.12$ & $10.30 \pm 0.10$ & $10.18 \pm 0.12$ & $10.38 \pm 0.16$ & $59.01 \pm 0.22$ \\ 
  \hline 
  Co$_{2}$Cr$_{2}$Fe$_{2}$Mn$_{2}$Ni$_{92}$ & $2.07 \pm 0.09$ & $2.10 \pm 0.08$ & $2.11 \pm 0.07$ & $2.07 \pm 0.11$ & $91.66 \pm 0.16$ \\ 
  \hline 
\end{tabular}
\end{table*}

\begin{figure*}
\begin{center}
  \includegraphics{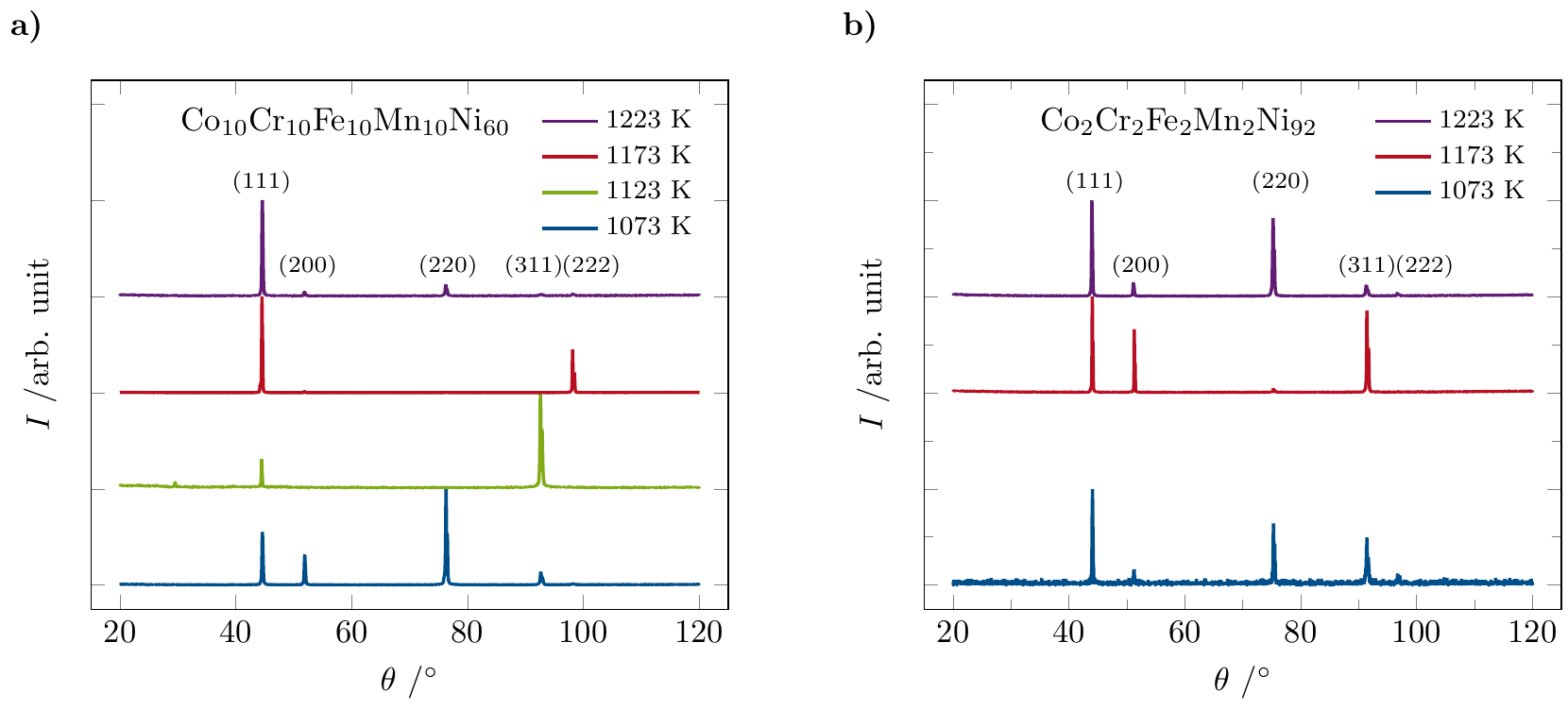}  
\end{center}
\caption{\footnotesize{X-ray diffraction patterns of (a)) Co$_{10}$Cr$_{10}$Fe$_{10}$Mn$_{10}$Ni$_{60}$ and (b)) Co$_{2}$Cr$_{2}$Fe$_{2}$Mn$_{2}$Ni$_{92}$ of the pre-annealed states. All patterns prove the formation of a single face-centred cubic phase.}
\label{fig:XRD}}
\end{figure*}

\begin{figure*}[tbp]
  \includegraphics{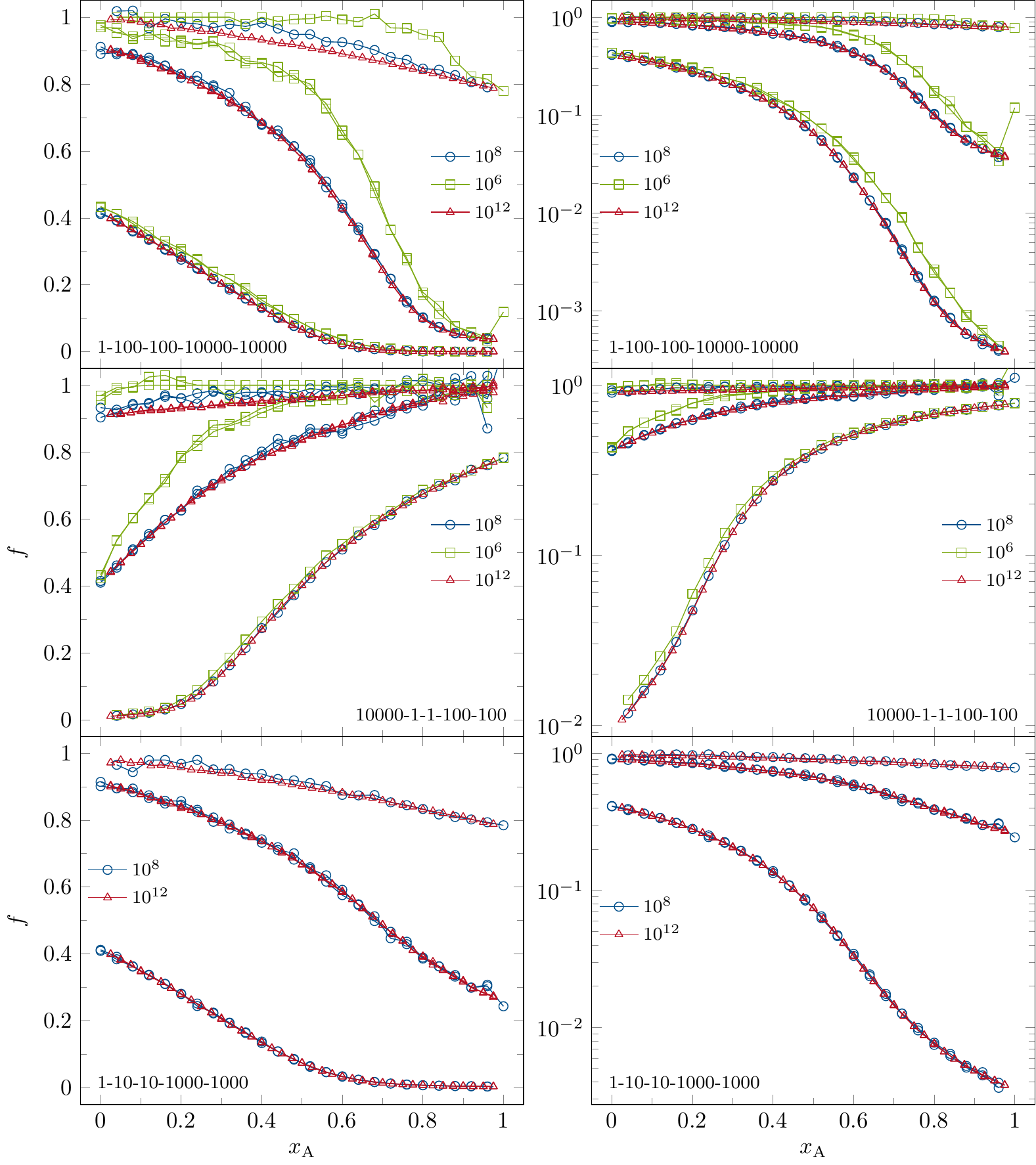}  
    \caption{Comparison of the correlation factor \(f\) using the KMC code developed for this publication compared to a conventional random alloy model implementation. The legend entry denotes the number of MC steps taken, \(10^{12}\) marks the conventional implementation of the random alloy model, while \(10^6\) and \(10^8\) data points were obtained from the new KMC implementation. The graphs on the left side show the deviations on a linear scale and the data on the right side are the same results plotted against a logarithmic scale.
    }\label{fig:Appendix_f}
\end{figure*}

Fig.~\ref{fig:Appendix_f} shows the tracer correlation factor obtained from the conventional random alloy model after \(10^{12}\) steps compared to the correlation factor determined with the new KMC code using a constant vacancy migration energy (\(10^6\) and \(10^8\) steps). Three different set jump rates are used as input \(1-100-100-10000-10000\), \(10000-1-1-100-100\), \(1-10-10-1000-1000\), for species A through E respectively. We find significant deviations between both codes after \(10^6\) steps but overall great agreement after \(10^8\) steps.

\begin{figure}[tbp]
\includegraphics{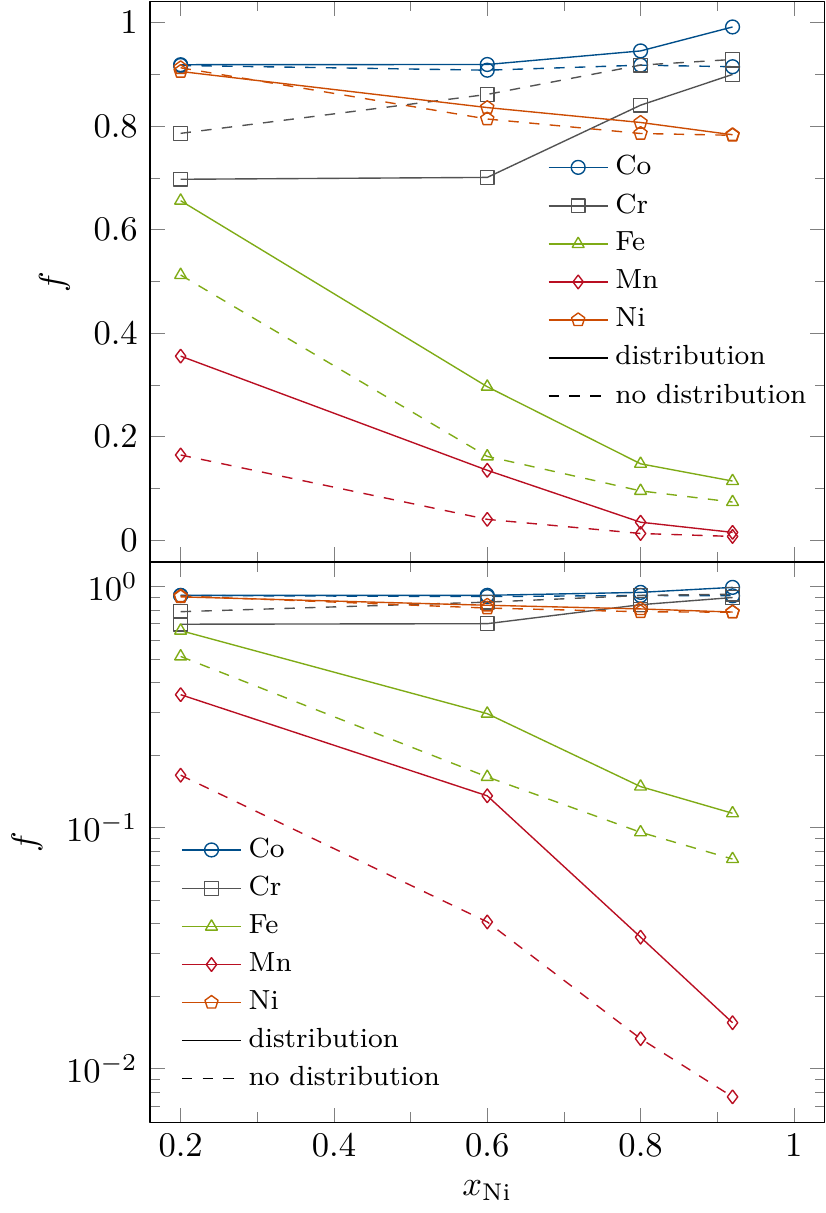}%
 \caption{Correlation factors obtained from the KMC simulation after \(10^8\) steps. Solid lines represent KMC results where the migration energy barrier are taken with a finite width (Tab.~\ref{tab:migrationBarrier}). Dashed lines show \(f\) with the a constant mean activation barrier but no distribution. This data corresponds to the random alloy model. In all cases \(f\) was sampled at \SI{1123}{\kelvin}.
 }\label{fig:fCantor_dist_nodist_appendix}
\end{figure}
\end{document}